\documentclass{elsart41}
\usepackage{graphics}
\usepackage{graphicx}
\usepackage{amssymb}
\begin{document}

\begin{frontmatter}



\vspace{-0.5cm}
\title{Probing the extended non-Fermi liquid regimes of MnSi and Fe}

%---- Don't remove this comment line! ----

\author[GE]{P. Pedrazzini\corauthref{Pedrazzini}}
%\ead{pedrazzi@mail.physics.unige.ch}
\ead{Pablo.Pedrazzini@physics.unige.ch}
\author[GE]{D. Jaccard}
\author[FR]{G. Lapertot}
\author[FR]{J. Flouquet}
\author[JA]{Y. Inada}
\author[JA]{H. Kohara}
\author[JA]{Y. Onuki}

%----------------------------------------------------------------------
% List of addresses
%
% If there is more than one address, list each using a separate 
% \address command using a label to link it to the respective author
% as described above
 
\address[GE]{DPMC, Universit\'e de Gen\`eve, Q. Ernest-Ansermet 24, 1211 Gen\`eve, Suisse}
\address[FR]{DRFMC, CEA Grenoble, 38054 Grenoble C\'edex 9, France}
\address[JA]{Grad. School of Science, Osaka University, Toyonaka, Osaka 560-0043, Japan}

\corauth[Pedrazzini]{Corresponding author. Tel: +41-22-379-6225}

\begin{abstract}

Recent studies show that the non-Fermi liquid (NFL) behavior of MnSi and Fe spans over an
unexpectedly broad pressure range, between the critical pressure $p_{\rm c}$ and around $2 p_{\rm
c}$. In order to determine the extension of their NFL regions, we analyze the evolution of the
resistivity $\rho(T)\sim A(p) T^n$ at higher pressures. We find that in MnSi the $n=3/2$ exponent
holds below $4.8\,{\rm GPa}\approx 3 p_{\rm c}$, but it increases above that pressure. At
$7.2\,$GPa we observe the low temperature Fermi liquid exponent $n=2$ whereas for $T>1.5\,$K,
$n=5/3$. Our measurements in Fe show that the NFL behavior $\rho\sim T^{5/3}$ extends at least up to
$30.5\,$GPa, above the entire superconducting (SC) region. In the studied pressure range, the onset
of the SC transition reduces by a factor $10$ down to $T_{\rm c}^{\rm onset}(30.5\,{\rm
GPa})=0.23\,$K, while the $A-$coefficient diminishes monotonically by around $50$\%.

\end{abstract}

%\vspace{-0.3cm}
\end{frontmatter}

% main text

Anomalous dependencies of thermodynamic and transport properties are observed in the pressure
studies of many metals close to magnetic instabilities. This {\em non-Fermi liquid} (NFL) behavior
is usually detected in a narrow region of the phase diagram, around a critical pressure, $p_{\rm
c}$, where the magnetic ordering temperature $T_{\rm ord}(p)\to 0$. However, a recent report
\cite{Doiron-Leyraud03} shows that in the weak helimagnet MnSi the electrical resistivity follows
$\rho(T)\sim T^{3/2}$ for pressures between $p_{\rm c}=1.46\,$GPa and around $2 p_{\rm c}$, at least
down to $0.5\,$K. This contrasts with the expected $\rho(T)$ in a nearly ferromagnetic metal (NFM):
a low temperature $\rho(T)\sim T^{2}$ and $\rho(T)\sim T^{5/3}$ above a characteristic $T^*$. 

Elemental iron displays, in its non-magnetic $\epsilon-$phase, non-Fermi liquid behavior with
striking characteristics.\cite{Holmes04} The NFL region emerges close to a strong first order
magnetic transition, followed by a structural one,\cite{Mathon04} and extends between $p_{{\rm
c}1}\approx 13\,$GPa and almost $2 p_{{\rm c}1}$.\cite{Holmes04} Furthermore, in this pressure
window Fe is an  unconventional superconductor (SC) with an onset at $T_{\rm c}^{\rm
onset}\approx 2\,$K and the resistivity $\rho(T)\sim T^{5/3}$ suggests the effect of ferromagnetic
(FM) spin fluctuations instead of the predicted antiferromagnetic ones.\cite{Holmes04} 

We have extended previous works by measuring four-probe electrical resistivity of single crystalline
MnSi and a Fe-whisker to higher pressures. The clamped Bridgman anvil cell technique was employed up
to $7.2\,$GPa and $30.5\,$GPa for MnSi and Fe, respectively, using steatite as pressure transmitting
medium. 

%%%%%%%%%%%%%%%%%%%%%%%%%%%%%%%%%%%%%%%%%%%%%%%%%%%%%%%%%%%%
%% MnSi
%%%%%%%%%%%%%%%%%%%%%%%%%%%%%%%%%%%%%%%%%%%%%%%%%%%%%%%%%%%%

%
%
\begin{figure}[ttt]
\begin{center}
\vspace{-0.6cm}
\includegraphics[width=0.385\textwidth]{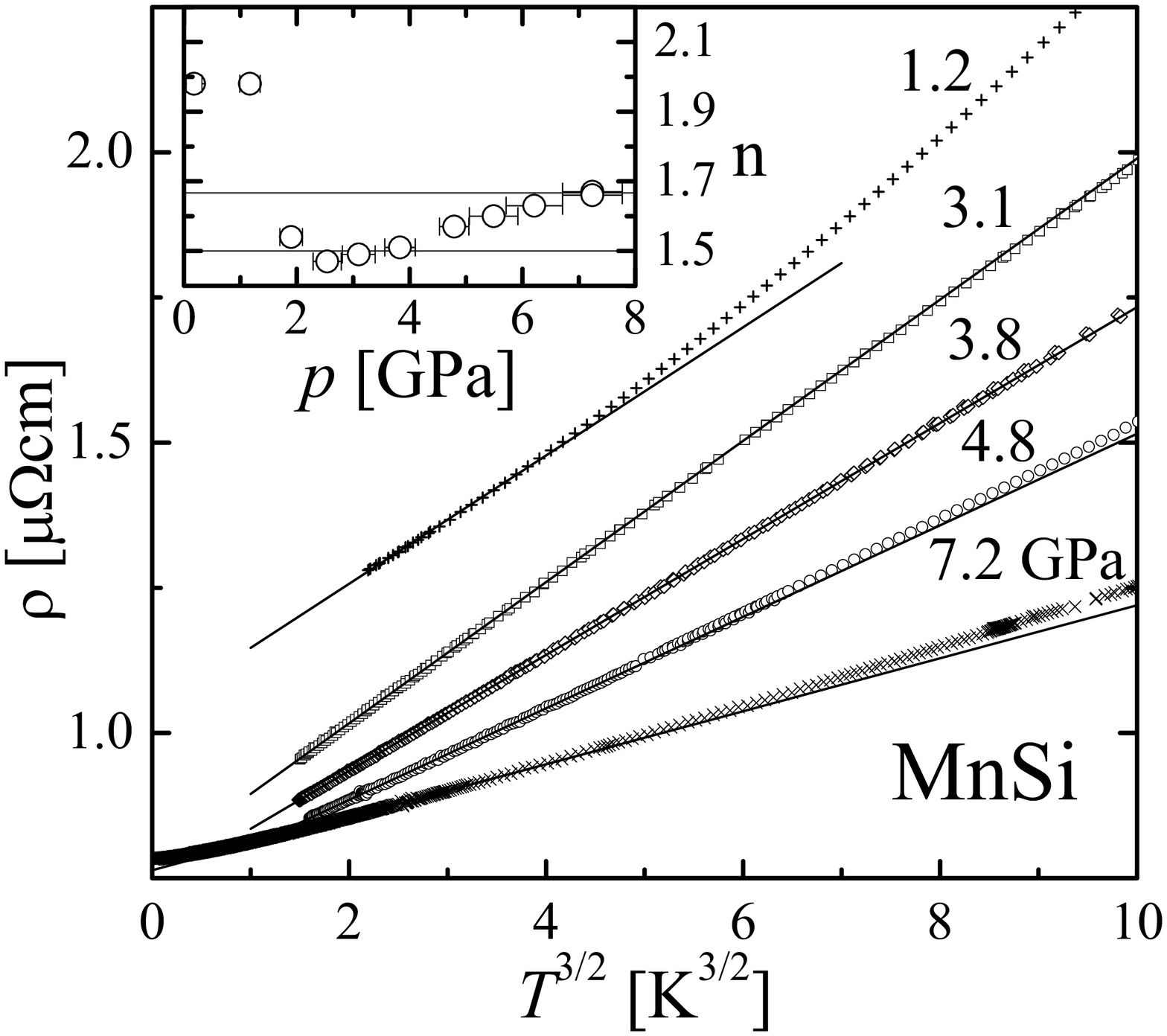}
\end{center}
%\vspace{-0.2cm}
\caption{Low temperature resistivity $\rho$ vs.~$T^{3/2}$ of MnSi at five chosen pressures. The
straight lines are fittings that stress the deviation from a $\rho(T)\sim T^{3/2}$ regime. Inset:
pressure-evolution of the $n-$exponent, as results from fitting our data to $\rho(T)=\rho_0+A T^n$
between $1.3$ and $4\,$K.}
\label{fig:figMnSi}
\end{figure}

In the case of MnSi, our main goal was to determine the maximum pressure up to which the abnormal
exponent $n=3/2$ holds. Low temperature resistivity data is presented in Fig.~\ref{fig:figMnSi} in a
$\rho$ vs.~$T^{3/2}$ representation. In addition, the inset of this figure depicts the evolution of
$n(p)$ as obtained from fitting our data to $\rho(T)=\rho_0+A T^n$ between $1.3$ and $4\,$K. For
$p=1.2\,$GPa, well below $T_{\rm ord}$, the resistivity follows $\rho(T)\sim T^2$. Indeed, at
$1.2\,$GPa, the resistivity data deviates upward from a straight line, suggesting a $n>3/2$
exponent. Such a deviation is not present in the data for pressures between $p_{\rm c}$ and
$4.8\,$GPa, indicating that $n\approx 3/2$ holds in this temperature range for $p_{\rm c}<p<3 p_{\rm
c}$. Above $3 p_{\rm c}$, $n(p)$ increases smoothly towards $n\approx 5/3$, see the figure inset.
Resistivity measurements performed down to $50\,$mK suggest that at $p=7.2\,$GPa the conventional
picture for a NFM is recovered: we find $n=2.0$ for $50\,{\rm mK}<T<1.3\,$K and the quoted $n=1.7$
up to $4\,$K, with an estimated  $T^*(p\approx 5 p_{\rm c})\sim 1.5\,$K. The coefficient $A=0.029
\mu \Omega\,{\rm cm\,K}^{-2}$ we observe for the $\rho(T)\sim A T^2$ behavior at $p=7.2\,$GPa is
within $7$\% of the value at $p=0.18\,$GPa, the lowest measured pressure. 

%%%%%%%%%%%%%%%%%%%%%%%%%%%%%%%%%%%%%%%%%%%%%%%%%%%%%%%%%%%%
%% Fe
%%%%%%%%%%%%%%%%%%%%%%%%%%%%%%%%%%%%%%%%%%%%%%%%%%%%%%%%%%%%

\begin{figure}[ttt]
\vspace{-0.6cm}
\begin{center}
\includegraphics[width=0.40\textwidth]{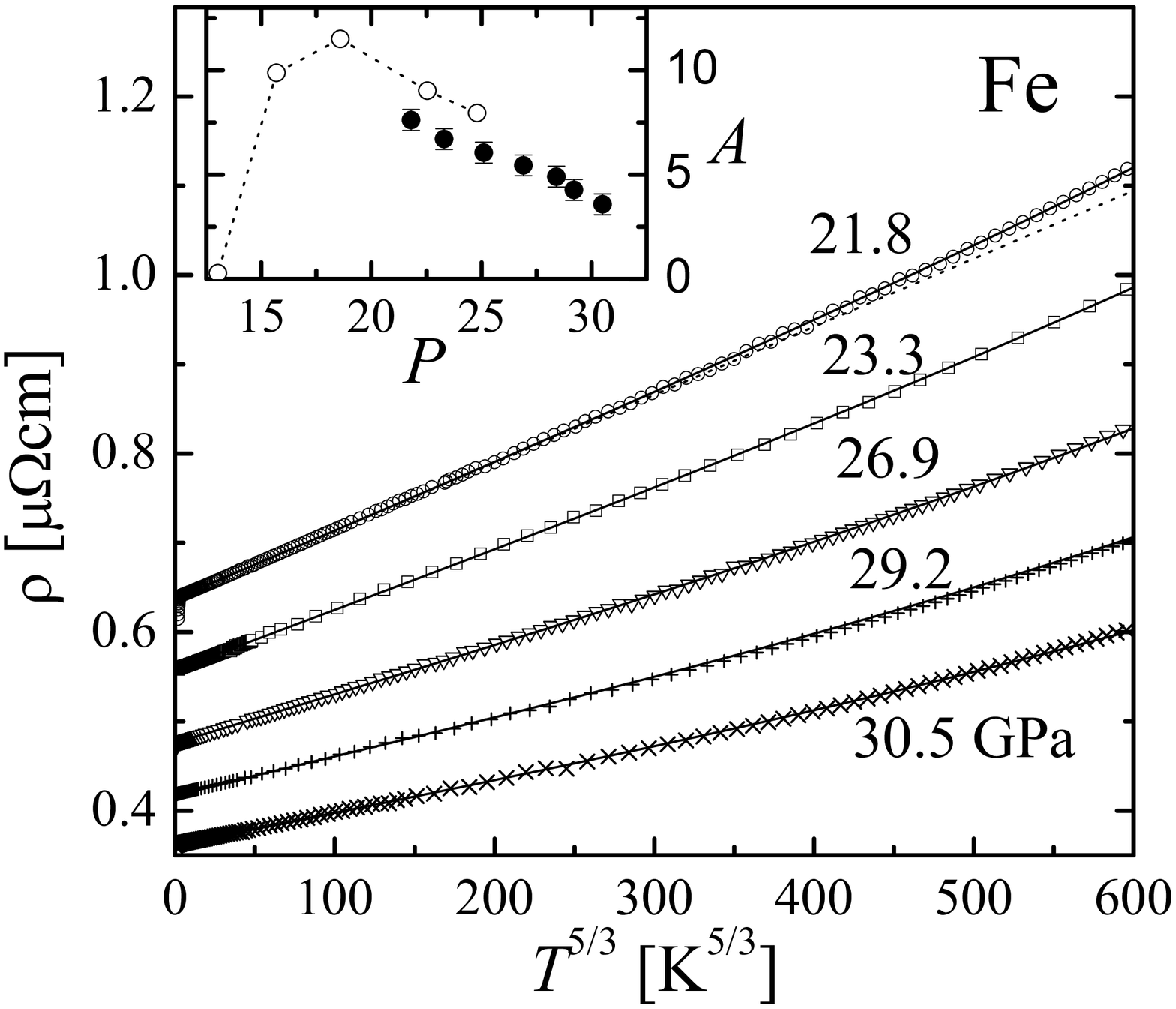} 
\end{center} 
%\vspace{-0.2cm}
\caption{Resistivity of iron for pressures between $21.8\,$GPa and $30.5\,$GPa in a $\rho$
vs.~$T^{5/3}$ representation. The dotted-line is the fit for the $p=21.8\,$GPa data to $\rho\sim
T^{5/3}$, while the full lines are fittings including a phonon contribution. Inset: evolution of the
$A-$coefficient (in units of $10^{-4} \mu \Omega\,{\rm cm\,K}^{-5/3}$); our data ($\bullet$) is
compared with the data of reference \cite{Holmes04} ($\circ$).} 
\label{fig:figFe}
\end{figure}

In Fig.~\ref{fig:figFe} we plot $\rho(T)$ of Fe as a function of $T^{5/3}$. Between $21.8\,$GPa
and $30.5\,$GPa, the normal state resistivity follows $\rho(T)\sim T^{5/3}$ in a broad temperature
range, as previously observed at lower pressures.\cite{Holmes04} The data departs from this simple
dependence above $30\,$K, compare in Fig.~\ref{fig:figFe} the dot-line with the data for
$p=21.8\,$GPa. We ascribe this departure to the extra $\rho(T)\sim T^5$ contribution due to
electron-phonon scattering, included in the solid-lines fits thru the data. If we restrict the fitting-range to low temperatures, between $T_{\rm c}^{\rm onset}(p)$ and $5\,$K, we find an exponent that scatters around $n\approx 1.75$ up to $p\approx 27\,$GPa and then diminishes down to $n\approx 1.4$ at $30.5\,$GPa.  

The $A(p)$ coefficient of Fe follows the remarkable evolution depicted in the inset of
Fig.~\ref{fig:figFe}: it increases abruptly at $p_{{\rm c}1}$ and tracks $T_{\rm c}^{\rm onset}(p)$
above that pressure.\cite{Holmes04} Our data extends the correlation between $T_{\rm c}(p)$ and
$A(p)$ almost up to $p_{\rm c2}\approx 31\,$GPa, where $T_{\rm c}^{\rm onset}(p)\to 0$. However, for
$21.8\,{\rm GPa}\le p\le 30.5\,{\rm GPa}$ we observe a reduction of $T_{\rm c}(p)$ in a factor $10$,
down to $T_{\rm c}^{\rm onset}(30.5\,{\rm GPa})=0.23\,$K, while $A(p)$ decreases by only $50$\%.
Within the SC scenario proposed for Fe,\cite{Holmes04} the strong suppression of
$T_{\rm c}^{\rm onset}(p)$ could be related to a collapse of the pairing mechanism, indicating a
threshold value of $A$ for the observation of SC. A further possibility is that a reduction in
$A(p)$, and thus in $T_{\rm c}^{\rm onset}(p)$, would imply a growing coherence length $\xi\sim
1/T_{\rm c}^{\rm onset}$. Considering that the mean free path $l$ is not changing much, as $p$
increases it is not possible to satisfy the clean-limit condition for SC in Fe, $l\gg \xi$.  

% 
% 
%\begin{figure}[!ht] 

The change of the magnetic state in MnSi is also reflected in the evolution of $A(p)$, which has a
strong increase when $p\to p_{\rm c}^{\pm}$. This effect is attributed to stron\-ger fluctuations in
the magnetization.\cite{Thessieu} As $p$ increases above $p_{\rm c}$, $A(p)$ diminishes smoothly,
see the slope of the full lines in Fig.~\ref{fig:figMnSi}. Although it is far from being proved, the
smooth evolution of $A(p)$ suggests that no phase-transition-line is crossed when the exponent \nolinebreak evolves from $n(p<3 p_{\rm c})=3/2$ to $n(p>3 p_{\rm c})=5/3$.

In this work we present further evidence for an {\em extended} NFL region in MnSi and Fe. In the
case of MnSi, we show that the abnormal $n=3/2$ exponent holds up to $3 p_{\rm c}$, but a
``conventional'' NFM is recovered at $7.2\,$GPa. The relevant question concerning this material is
how $T^*(p)$ evolves; higher pressure experiments or $\rho(T)$ measurements with increased
sensitivity could partially answer this. The case of Fe is similar to that of MnSi in the sense that
it is an extended NFL with striking characteristics. We have also pointed out that SC
is suppressed at around $31\,$GPa despite the strong electronic correlation measured by $A$. In this
case, an open question is the influence that the first-order magnetic/structural transition may have
for $p>p_{\rm c1}$. Further pressure experiments on other $3d-$FM, in especial on {\em pure} systems
like the ones studied here, may give important clues about relevant scenarios.

%%%%%%%%%%%%%%%%%%%%%%%%%%%%%%%%%%%%%%%%%%%%%%%%%%%%%%%%%%%%%%%%%%%%%
% \vspace{-0.5cm}

\end{document}